\documentclass[11pt]{article}

\usepackage[final]{acl}
\usepackage{float}
\usepackage{times}
\usepackage{latexsym}
\usepackage{subfig} 
\usepackage[table]{xcolor}
\usepackage{fontawesome} 
\usepackage{pifont}
\usepackage{natbib}
\usepackage{amsmath}
\usepackage{amssymb}
\usepackage{booktabs}
\usepackage[most]{tcolorbox}
\tcbuselibrary{breakable,skins,listings}
\usepackage{multirow}
\usepackage{siunitx}
\usepackage{booktabs}

\newcommand{\systemtag}{\textcolor{PromptBlue}{\textbf{\textit{System Prompt}}}\ }
\newcommand{\prompttag}{\textcolor{PromptRed}{\textbf{\textit{User Prompt}}}\ }
\newcommand{\chg}[1]{\scriptsize(#1)}

\usepackage[T1]{fontenc}

\usepackage[utf8]{inputenc}

\usepackage{microtype}

\usepackage{inconsolata}

\usepackage{graphicx}

\title{Interpretable All-Type Audio Deepfake Detection with Audio LLMs via Frequency–Time Reinforcement Learning}

\author{
Yuankun Xie\textsuperscript{1,2},
Xiaoxuan Guo\textsuperscript{1,2},
Jiayi Zhou\textsuperscript{2},
Tao Wang\textsuperscript{2},
Jian Liu\textsuperscript{2}$^{*}$, \\
\bf Ruibo Fu\textsuperscript{3},
\bf Xiaopeng Wang\textsuperscript{3},
\bf Haonan Cheng\textsuperscript{1},
\bf Long Ye\textsuperscript{1}\thanks{Corresponding author} \\
\textsuperscript{1}Communication University of China, Beijing, China \\
\textsuperscript{2}Machine Intelligence, Ant Group, Shanghai, China \\
\textsuperscript{3}Institute of Automation, Chinese Academy of Sciences, Beijing, China
\\
 \small{xieyuankun@cuc.edu.cn}
}

\begin{document}
\maketitle
\begin{abstract}
Recent advances in audio large language models (ALLMs) have made high-quality synthetic audio widely accessible, increasing the risk of malicious audio deepfakes across speech, environmental sounds, singing voice, and music. Real-world audio deepfake detection (ADD) therefore requires all-type detectors that generalize across heterogeneous audio and provide interpretable decisions. Given the strong multi-task generalization ability of ALLMs, we first investigate their performance on all-type ADD under both supervised fine-tuning (SFT) and reinforcement fine-tuning (RFT). However, SFT using only binary real/fake labels tends to reduce the model to a black-box classifier, sacrificing interpretability. Meanwhile, vanilla RFT under sparse supervision is prone to reward hacking and can produce hallucinated, ungrounded rationales. To address this, we propose an automatic annotation and polishing pipeline that constructs Frequency-Time structured chain-of-thought (CoT) rationales, producing \(\sim\)340K cold-start demonstrations. Building on CoT data, we propose Frequency Time-Group Relative Policy Optimization (FT-GRPO), a two-stage training paradigm that cold-starts ALLMs with SFT and then applies GRPO under rule-based frequency-time constraints. Experiments demonstrate that FT-GRPO achieves state-of-the-art performance on all-type ADD while producing interpretable, FT-grounded rationales. The data and code are available online\footnote{https://xieyuankun.github.io/FT-GRPO}.
\end{abstract}

\section{Introduction}
\begin{figure}[!tb]
	\centering
	\subfloat{\includegraphics[width=2.9in]{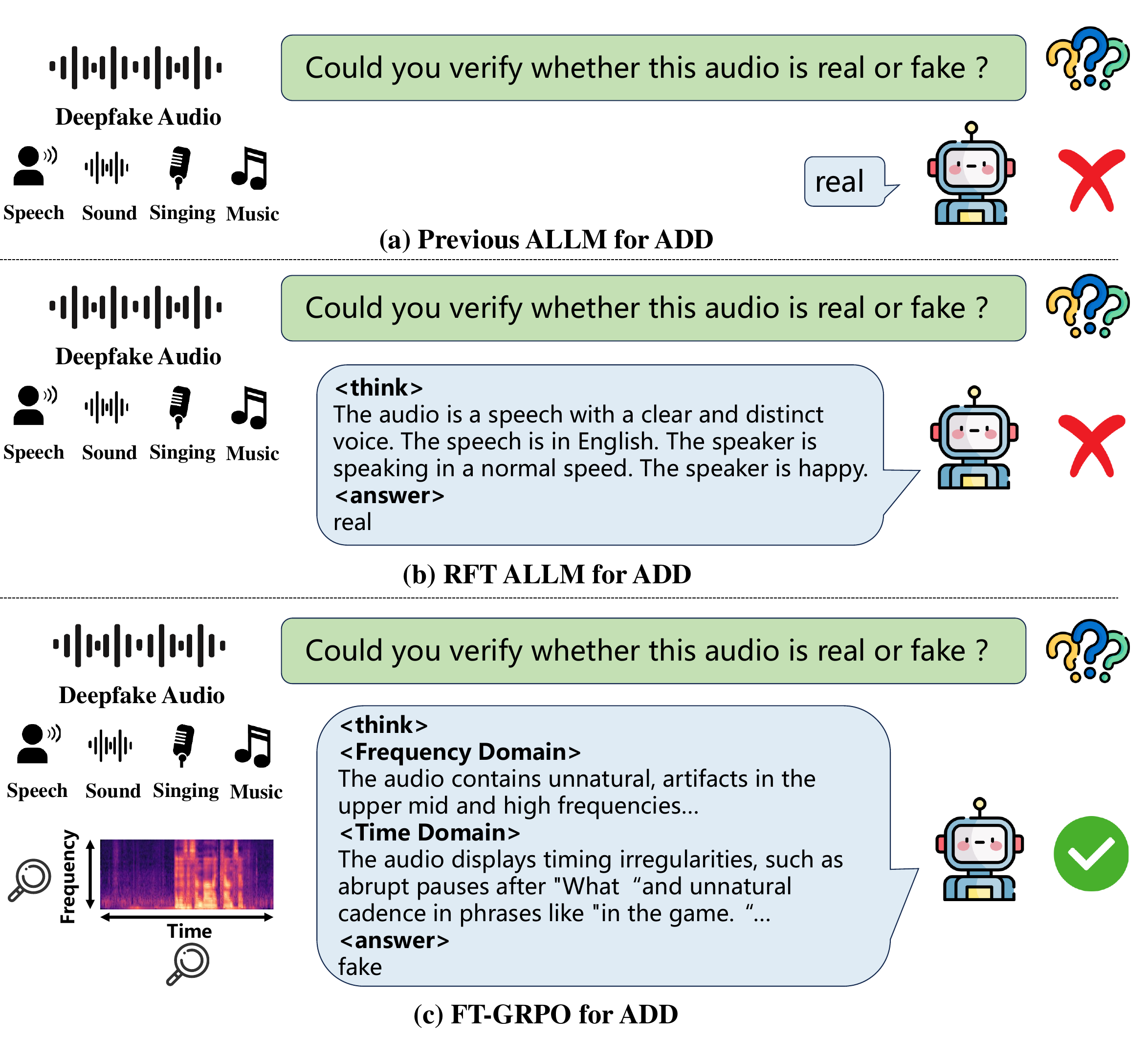}}
	\hfil
	\caption{Three ALLM-based CMs for the all-type ADD task: (a) label-only SFT, (b) RFT, and (c) our proposed FT-GRPO.}
	\label{fig:motivation} 
\end{figure}
With recent advances in audio large language model (ALLM), generating high-quality synthetic audio, including deepfake speech, environmental sounds, singing voices, and music, has become increasingly accessible. Such deepfake content poses growing risks across media and entertainment, cybersecurity, and political communication.

Research on audio deepfake detection (ADD) has expanded steadily in recent years. Early efforts primarily targeted deepfake speech, leading to a broad range of countermeasures (CMs) developed under the ASVspoof challenges \cite{todisco19_interspeech, liu2023asvspoof, wang2024asvspoof5}. Recently, the scope of ADD has moved beyond speech to cover other audio domains, including singing voice deepfake detection \cite{zhang2024svdd, xie2024fsd, zang2024ctrsvdd}, environmental sound deepfake detection \cite{yin25_interspeech, xie2024fakesound}, and music deepfake detection \cite{comanducci2024fakemusiccaps}. However, most existing studies are confined to a single audio type. In real-world ADD scenarios, detectors are expected to determine authenticity for arbitrary audio types. More recently, several works have investigated cross-type settings involving two or more audio domains~\cite{gohari2025audio, xie2025detect}; nevertheless, these efforts still lack both strong cross-type generalization and interpretable decision-making.

ALLMs inherit substantial pre-training knowledge and strong comprehension capabilities. Recent studies on the trustworthiness of ALLMs~\cite{li2025audiotrust,yang2025towards} have begun to touch upon deepfake detection, but only at a preliminary stage. Gu et al.~\cite{gu2025allm4add} provide the first systematic investigation of employing ALLMs for ADD. They show that zero-shot prompting often fails because ALLMs do not reliably internalize the authenticity judgment objective, leading to poor performance; supervised fine-tuning (SFT) with binary real/fake labels improves accuracy, yet it typically reduces the model to a black-box classifier and may suppress its inherent reasoning and explanation abilities. Moreover, existing ALLM-based ADD studies are largely confined to the speech domain. 

These limitations highlight a central challenge for real-world deployment: \emph{can we train ALLMs to solve the all-type ADD task with both strong cross-type generalization and interpretable decision rationales?} Reinforcement learning (RL) offers a promising direction toward this goal. In the vision domain, Visual-RFT \cite{liu2025visual} demonstrates that RL-based fine-tuning can outperform SFT while better preserving the interpretability of the underlying language model. Motivated by this evidence, we investigate whether RL can similarly benefit ALLMs for all-type ADD.

To this end, we revisit all-type ADD from the perspective of ALLMs and reinforcement fine-tuning (RFT). 
Specifically, we first establish an ALLM-based CM benchmark for all-type ADD setting by conducting SFT on representative ALLMs, in order to quantify their current performance and cross-type generalization. 
To further improve both interpretability and generalizability, we then explore direct RFT. 
However, as shown in Figure~\ref{fig:motivation}, RFT alone often fails to produce reliable, evidence-grounded rationales, as its supervision is typically sparse and is often restricted to format and final-label constraints, providing little guidance for reasoning and encouraging reward hacking or degenerate explanations~\cite{zhou2025r1}.

Inspired by the success of DeepSeek-R1’s RL framework with an SFT cold start~\cite{guo2025deepseek}, and by recent evidence that SFT-based cold starts strengthen RL for multimodal LLMs~\cite{chen2025advancing,wei2025advancing}, we first construct frequency–time (FT) structured chain-of-thought (CoT) rationales for existing all-type ADD datasets via an automatic annotation-and-polishing pipeline. This provides high-quality demonstrations that teach ALLMs a principled reasoning schema. Although other cues may help, we focus on frequency and time as concise, broadly applicable views that have been shown effective for capturing discriminative spoofing artifacts in prior lightweight CMs~\cite{tak2021end1,jung2022aasist}. Building on this cold start, we further optimize the model with RL. Specifically, we propose Frequency-Time Group Relative Policy Optimization (FT-GRPO), a two-stage training paradigm that combines SFT initialization with GRPO under rule-based frequency and time domain constraints. We also effectively leverage annotation-flagged non-think mismatch samples, where the rationale conflicts with the label, to improve detection accuracy and interpretability.

We summarize the contributions of this work as follow:
\begin{itemize}
  \item To quantify the cross-type generalization of ALLM-based countermeasures for all-type ADD task, we first establish a benchmark by SFT representative ALLMs.
  \item To improve interpretability of ALLMs, we propose an automatic annotation-and-polishing pipeline that constructs CoT rationales for four public datasets, yielding $\sim$340K cold-start demonstrations; to our knowledge, this is the first rationale annotated resource in the ADD literature.
  \item We propose FT-GRPO, a two-stage paradigm that combines SFT cold start with GRPO under rule-based FT reasoning constraints and explicitly leverages non-think samples to improve accuracy and interpretability.
  \item FT-GRPO achieves state-of-the-art (SOTA) results with a 3B model trained only on speech data, reaching 99.75\% accuracy on the ASVspoof2019LA evaluation set while producing interpretable rationales; co-training on all types further achieves 90.10\% average accuracy across all test sets.
\end{itemize}

\section{Related Work}
\subsection{Audio Deepfake Detection}
In introducing the current work in the field of ADD, we categorize it based on types: speech, sound, singing voice, and music.

\textbf{Speech}. Speech deepfake detection has been extensively studied, largely driven by the ASVspoof challenges. Representative countermeasures include AASIST \cite{jung2022aasist} and SSL-based pipelines that combine XLSR with AASIST \cite{tak2022automatic}. Subsequent work explores different SSL backbones \cite{phukan2024heterogeneity,kheir2025comprehensive}, layer utilization \cite{zhang2024audio,wang2025mixture,pan2024attentive}, and robustness \cite{zhang2025i,kawa2023defense}.

\textbf{Other types (sound, singing, music)}. Compared to the work on speech deepfake detection, research on other types is still primarily at the dataset construction stage, with fewer methodological explorations. For sound, recently, the Environmental Sound Deepfake Detection (ESDD) Challenge \cite{yin25_interspeech} has greatly enriched research on deepfake sound detection. ESDD covers a wide range of ALLM‑based text-to-audio (TTA) and audio-to-audio (ATA) synthesis methods, and the current SOTA solutions typically rely on sound SSL features such as SSLAM \cite{guo2025envsslam}. For singing voice, SVDD \cite{zhang2024svdd} motivates methods that combine XLSR with music-oriented SSL models such as MERT \cite{li2024mert} and WavLM \cite{chen2022wavlm} \cite{10832226,zhang2024xwsb,chen2024singing}. For music, FakeMusicCaps \cite{comanducci2024fakemusiccaps} supports synthetic-music detection, though methodological studies remain limited \cite{li2024detecting,wei2025voices}.

\textbf{Cross-type}. A few works study transfer across audio types, e.g., between speech and singing voice \cite{gohari2025audio} or from speech to music \cite{li2024audio}. Xie et al. \cite{xie2025detect} further establish an SSL-based benchmark for all-type ADD and propose wavelet prompt tuning to improve generalization.

\subsection{ALLM for Audio Deepfake Detection}
ALLMs possess rich pre‑training knowledge and strong comprehension capabilities, and have been successfully applied to a variety of tasks such as audio reasoning and captioning \cite{zhifei-etal-2025-audio,tang2024extending,ma2025mmar}, speech recognition and generation \cite{fathullah2024prompting,zhang2024speechgpt,yan2025ming}, and quality assessment \cite{wang2025speechllm,wang2025qualispeech}, among others. Recent work on the trustworthiness of ALLMs \cite{li2025audiotrust,yang2025towards} has also begun to touch on deepfake detection, but only at a preliminary level. Gu et al. \cite{gu2025allm4add} present the first systematic study of using ALLMs for the ADD task. They perform SFT on Qwen-Audio for binary real/fake classification and conduct a detailed investigation of ALLM performance from prompt design to zero-shot, few-shot, and hyperparameter settings, achieving an ACC of 99.40\% on the 19LA benchmark, surpassing previous small audio models. However, restricting the output to a simple real/fake label severely limits the interpretability of ALLMs.

\section{Method} 
\begin{figure*}[!t]
	\centering
	\subfloat{\includegraphics[width=6in]{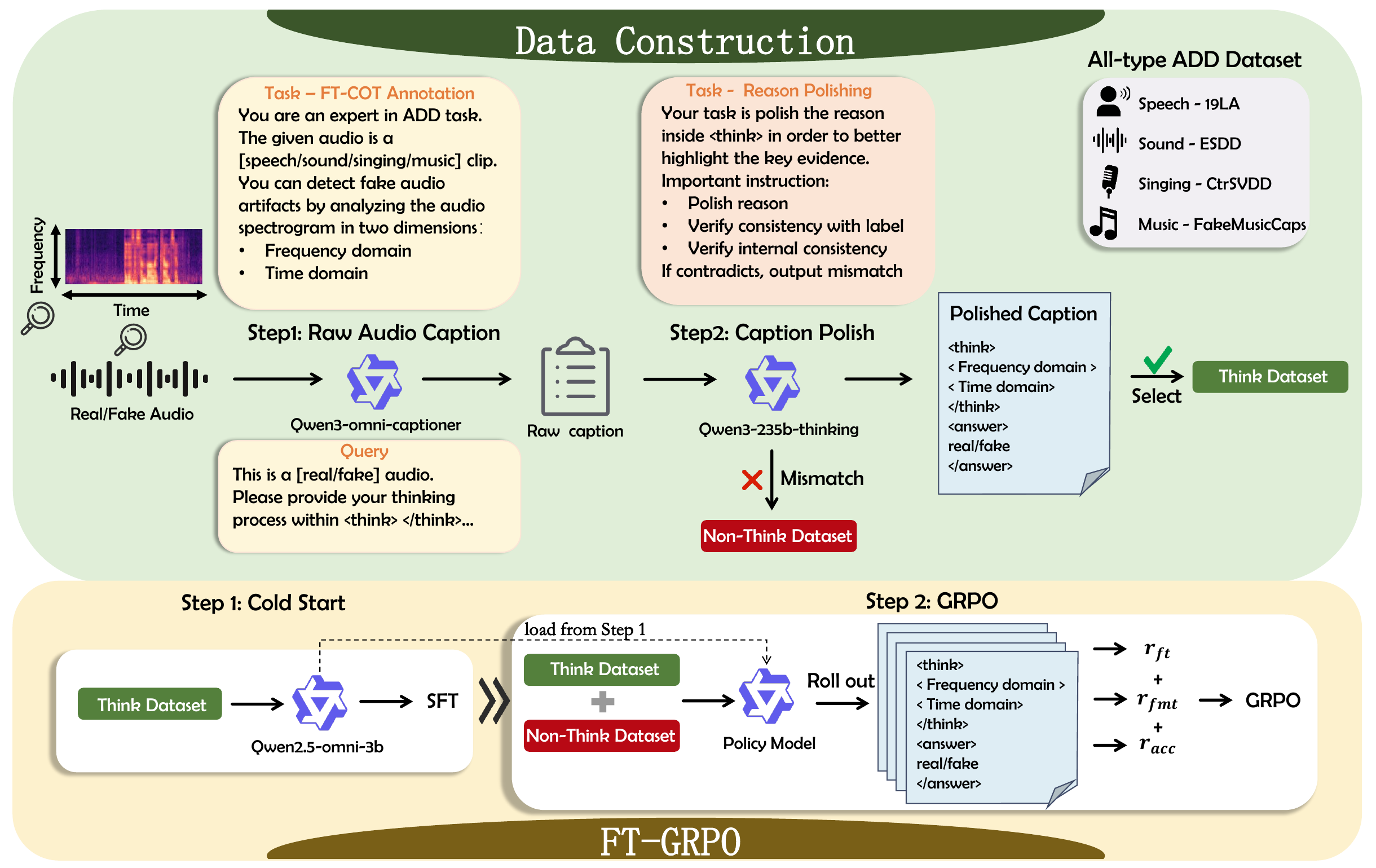}}
	\hfil
	\caption{Pipeline overview. Top: data construction via Step-1 raw audio caption and Step-2 caption polish. Bottom: FT-GRPO training, consisting of Step-1 SFT cold start and Step-2 GRPO.}
	\label{fig:pipeline} 
\end{figure*}
Figure~\ref{fig:pipeline} overviews our approach.
We first construct FT rationales via a two-step automatic captioning and polishing pipeline, producing a \textit{think} subset with valid rationales and a \textit{non-think} subset.
We then propose FT-GRPO, a two-stage training paradigm that (i) cold-starts an ALLM with SFT on the think subset to learn the FT reasoning schema, and (ii) applies GRPO with composite rewards to improve detection accuracy while encouraging complete FT-grounded rationales.
\subsection{Data Construction} 
Contemporary ALLMs still exhibit limited reasoning and explanatory abilities. Therefore, we first construct rationale-annotated data to elicit their reasoning behaviors and improve task understanding. 
Accordingly, for an sample \(S=\{a,y\}\) from a ADD dataset \(D\), where \(a\) denotes the audio clip and \(y\in\{\text{real},\text{fake}\}\) is the ground-truth label, we adopt a two-step caption generation pipeline.

\textbf{Step 1: Raw Audio Caption.}
We first use the SOTA audio captioning model Qwen3-Omni-Captioner\footnote{https://huggingface.co/Qwen/Qwen3-Omni-30B-A3B-Captioner}, which provides fine-grained audio understanding and is designed to produce accurate, comprehensive descriptions in complex and diverse acoustic scenarios.
Without additional prompting, it can parse and describe a wide range of audio content, including speech, environmental sounds, music, and cinematic sound effects, making it well suited for our all-type ADD task.

In step~1, each audio sample \(S\) is fed into the captioner.
Given the label \(y\), we instruct the model to reason from both the frequency-domain and time-domain perspectives, writing the reasoning process in \texttt{<think>} and the label \(y\) in \texttt{<answer>}, thereby producing a raw caption \(R_c\).
However, \(R_c\) exhibits several issues:
(i) the captions are often overly verbose and contain substantial information irrelevant to authenticity judgment;
(ii) some outputs are internally inconsistent, with \texttt{<think>} contradicting \texttt{<answer>} (e.g., reasoning indicates real while the answer is fake);
and (iii) the predicted \texttt{<answer>} of some outputs still contradicts the provided label \(y\) despite being explicitly given.

\textbf{Step 2: Caption Polish.}
To address these issues, we employ the LLM Qwen3-235B\footnote{https://huggingface.co/Qwen/Qwen3-235B-A22B} to polish each \(R_c\).
For issue (i), Qwen3-235B rewrites the description to be concise, removes content unrelated to authenticity cues, and enforces the required output format.
For issues (ii) and (iii), Qwen3-235B outputs a \texttt{<mismatch>} flag when it detects contradictions or cannot reliably reconcile the raw caption with the given label.
After this step, we obtain a polished caption \(P_c\) for most samples \(S\).

We treat samples flagged as \texttt{<mismatch>} as non-think cases that current ALLMs fail to handle robustly.
For these non-think cases, we retain only their \texttt{<answer>} and use them as reinforcement-learning data \(D_n\).
For samples with valid \(P_c\), we use them both as (i) cold-start data for FT-GRPO Step-1 SFT \(D_{\mathrm{think}}\) and (ii) training data for Step-2 GRPO.
Accordingly, Step-2 GRPO is trained on \(D_{\mathrm{r}} = D_{\mathrm{think}} \cup D_{\mathrm{non\_think}}\).
This design is motivated by the observation that SFT on data without reliable \texttt{<think>} annotations can negatively affect training; therefore, we use only correctly polished, format-consistent samples as high-quality data for cold-start initialization.

\subsection{FT-GRPO}
After the data construction process, we introduce \textbf{FT-GRPO}, our proposed training paradigm that consists of two steps: (i) Step 1: SFT cold start and (ii) Step 2: GRPO.

\textbf{Step 1: SFT Cold Start.}
In this step, we conduct SFT on \(D_{\mathrm{think}}\) (all samples with polished reasonable \texttt{<think>} annotations).
This enables the ALLM to learn how to reason for ADD task from both the frequency-domain and time-domain perspectives, and to generate outputs in the desired schema.

Let \(\pi_{\theta}\) be an ALLM with parameters \(\theta\).
For each SFT example \((a_i, P_{c,i}) \in D_{\mathrm{think}}\), the target \(P_{c,i}\) is a structured sequence containing \texttt{<think>} and \texttt{<answer>}.
We train \(\pi_\theta\) by maximizing the conditional likelihood of \(P_c\) given the audio input \(a\):
{\small
\begin{equation}
\begin{aligned}
\mathcal{L}_{\mathrm{SFT}}(\theta)
&=
-\mathbb{E}_{(a,P_c)\sim D_{\mathrm{think}}}
\left[
\sum_{t=1}^{|P_c|}
\log \pi_{\theta}\!\left(x_t \mid x_{<t}, a\right)
\right].
\end{aligned}
\label{eq:sft}
\end{equation}
}
where \(P_c=(x_1,\dots,x_{|P_c|})\) and \(x_{<t}\) denotes the prefix tokens \((x_1,\dots,x_{t-1})\).

\textbf{Step 2: GRPO Optimization.}
During reinforcement learning, we further optimize the Step-1 SFT model as policy using GRPO with a composite reward that simultaneously encourages prediction correctness, enforces well-formed outputs, and promotes complete frequency--time (FT) reasoning.
Given an audio clip \(a\) with ground-truth label \(y\in\{\text{real},\text{fake}\}\), the policy model generates an output
\(o=\{\tau, y_{\text{pred}}\}\), where \(\tau\) denotes the reasoning path inside \texttt{<think>} and \(y_{\text{pred}}\) is the predicted label inside \texttt{<answer>}.

\paragraph{Accuracy Reward.}
We define a sparse accuracy reward to encourage correct classification:
\begin{equation}
r_{\mathrm{acc}}=
\begin{cases}
1, & y_{\text{pred}} = y,\\
0, & \text{otherwise}.
\end{cases}
\label{eq:racc}
\end{equation}
\paragraph{Format Reward.}
To stabilize training, we enforce a fixed output schema and define a binary format reward:
\begin{equation}
r_{\mathrm{fmt}}=\mathbb{I}\!\left(\mathcal{F}(o)\right),
\label{eq:rfmt}
\end{equation}
where \(\mathcal{F}(o)\) is true if the reasoning \(\tau\) is wrapped by \texttt{<think>} and the predicted label \(y_{\text{pred}}\) is wrapped by \texttt{<answer>}.

\paragraph{FT Reasoning-Path Reward.}
Since \(r_{\mathrm{acc}}\) and \(r_{\mathrm{fmt}}\) are sparse and do not directly shape \(\tau\), we introduce an FT reasoning reward \(r_{\mathrm{ft}}\) that encourages the model to provide evidence from both the frequency domain and the time domain.
Let \(\mathcal{A}=\{\textsc{FD},\textsc{TD}\}\) denote the two meta-steps with declaration tags
\(a_{\textsc{FD}}\!=\)\texttt{<Frequency Domain>} and \(a_{\textsc{TD}}\!=\)\texttt{<Time Domain>}.
We assign:
\begin{equation}
r_{\mathrm{ft}}=\frac{1}{2}\sum_{i\in\mathcal{A}}
\mathbb{I}\!\left(a_i\in\tau \ \land\ g(a_i)=\texttt{True}\right),
\label{eq:rft}
\end{equation}
where \(g(a_i)\) is a lightweight rule-based completeness check (the tag is present and followed by at least one complete sentence).
Thus \(r_{\mathrm{ft}}\in\{0,0.5,1.0\}\), indicating missing, partial, and complete FT reasoning, respectively.

\paragraph{Total Reward.}
The final reward for an output \(o\) is computed as
\begin{equation}
r = r_{\mathrm{acc}} + \alpha \, r_{\mathrm{fmt}} + \beta \, r_{\mathrm{ft}} .
\label{eq:rtotal}
\end{equation}

\paragraph{GRPO Objective.}
We employ GRPO to optimize the policy model.
For each training instance (audio query) \(a\), the current policy \(\pi_{\theta}\) samples a group of \(G\) responses
\(\{o_1,\ldots,o_G\}\).
We compute rewards \(\{r_1,\ldots,r_G\}\) using Eq.~\eqref{eq:rtotal}, and estimate the group-relative advantage for the \(i\)-th response as
\begin{equation}
A_i =
\frac{r_i - \mathrm{mean}(\{r_1,\ldots,r_G\})}
{\mathrm{std}(\{r_1,\ldots,r_G\})+\epsilon},
\label{eq:adv_grpo}
\end{equation}
where \(\mathrm{mean}(\cdot)\) and \(\mathrm{std}(\cdot)\) denote the mean and standard deviation within the group, and \(\epsilon\) is a small constant for numerical stability.
The policy model is then updated to increase the likelihood of responses with higher \(A_i\), while regularizing the update with a KL penalty to a reference policy \(\pi_{\mathrm{ref}}\) (the SFT snapshot), as in standard GRPO.

\section{Experiments}
\begin{table}[t]
	\centering
	\setlength{\tabcolsep}{2.5pt}
	\fontsize{8pt}{9pt}\selectfont
	\renewcommand{\arraystretch}{0.95}
	\begin{tabular}{c|c|ccc}
		\toprule
		Type &Source & Train & Dev & Eval \\
		\midrule
		Speech & 19LA & 25,380 & 24,844 & 71,237 \\
		Sound & ESDD & 139,055 & 39,710 &4,000 \\
		Singing & CtrSVDD & 84,404	&43,625	&92,769 \\
		Music & FakeMusicCaps &20,861 &6,058	&6,122 \\
		\midrule
		All & Combined Sources & 199,023 & 84,438 & 189,951 \\
		\bottomrule
	\end{tabular}
	\caption{All-type ADD dataset in terms of original training, development, and evaluation set.}
    \label{tab:dataset}
\end{table}

\subsection{Dataset}
\label{dataset}
To evaluate the capability of CMs for all-type ADD, appropriate dataset selection is essential. We prioritize datasets that are relatively clean and exclude partially spoofed cases, so that the evaluation focuses on deepfake artifacts rather than confounding factors such as background noise or mixed genuine--spoof segments. This controlled setting facilitates a clearer analysis of cross-type ADD generalization. Original dataset details are summarized in Table \ref{tab:dataset}.
For detailed information on the FT-GRPO training and validation data, as well as the partitioning of think and non-think samples, please refer to appendix section~\ref{sec:appendix_data}.

\begin{table*}[t]
\centering
\small
\setlength{\tabcolsep}{4pt}
\renewcommand{\arraystretch}{1.08}
\begin{tabular}{llccccc}
\toprule
\textbf{Type} & \textbf{ALLM} & \textbf{Speech} & \textbf{Sound} & \textbf{Singing} & \textbf{Music} & \textbf{AVG} \\
\midrule
Speech  & Qwen2-Audio-Chat-7B & 89.99 & 67.05 & \textbf{84.85} & 82.33 & 81.06 \\
Speech  & Qwen2.5-Omni-3B     & \textbf{99.04} & \textbf{73.31} & 66.29 & \textbf{91.16} & \textbf{82.45} \\
Speech  & Qwen2.5-Omni-7B     & 96.32 & 61.95 & 80.30 & 90.80 & 82.34 \\
\addlinespace
Sound   & Qwen2-Audio-Chat-7B & 32.24 & 62.60 & 85.40 & 14.88 & 48.78 \\
Sound   & Qwen2.5-Omni-3B     & \textbf{86.70} & \textbf{67.55} & 77.99 & \textbf{93.58} & \textbf{81.46} \\
Sound   & Qwen2.5-Omni-7B     & 77.23 & 64.32 & \textbf{85.68} & 90.36 & 79.40 \\
\addlinespace
Singing & Qwen2-Audio-Chat-7B & 49.01 & 52.22 & 79.53 & 65.39 & 61.54 \\
Singing & Qwen2.5-Omni-3B     & 39.78 & \textbf{67.13} & \textbf{83.97} & \textbf{73.39} & 66.07 \\
Singing & Qwen2.5-Omni-7B     & \textbf{73.29} & 47.55 & 81.50 & 70.81 & \textbf{68.29} \\
\addlinespace
Music   & Qwen2-Audio-Chat-7B & 54.49 & \textbf{74.42} & 37.10 & \textbf{88.96} & 63.74 \\
Music   & Qwen2.5-Omni-3B     & 83.72 & 73.40 & 23.06 & 77.34 & 64.38 \\
Music   & Qwen2.5-Omni-7B     & \textbf{89.28} & 70.57 & \textbf{83.41} & 88.47 & \textbf{82.93} \\
\bottomrule
\end{tabular}
\caption{ACC results (\%) of ALLM-based CM training and testing on all-type ADD dataset.}
\label{tab:alltype_results}
\vspace{-2mm}
\end{table*}

\textbf{Speech-19LA}. A widely adopted benchmark comprising 12,456 real and 108,978 fake samples generated by 11 TTS and 8 VC systems (A01--A19). Following the standard protocol, A01--A06 are used for training and A07--A19 for evaluation, ensuring that the spoofing systems in the evaluation set are unseen during training.

\textbf{Sound-ESDD}. ESDD \cite{yin25_interspeech} is the first challenge dedicated to environmental-sound deepfake detection. It covers seven ALM-based synthesis methods spanning both text-to-audio (TTA) and audio-to-audio (ATA) generation. The training and validation splits include methods G01--G04, while the test split contains previously unseen methods G05--G07.

\textbf{Singing Voice-CtrSVDD}. SVDD~\cite{zhang2024svdd} is the first challenge dedicated to singing-voice deepfake detection. We use its CtrSVDD subset, built from Mandarin and Japanese singing corpora and synthesized by 14 SVS/SVC systems. We follow the official split: A01--A08 for training and A09--A14 for testing.

\textbf{Music-FakeMusicCaps}. FakeMusicCaps~\cite{comanducci2024fakemusiccaps} is a dataset for deepfake music detection. Its real-domain audio comes from MusicCaps~\cite{agostinelli2023musiclm}, which contains 5.5k 10-second music clips from AudioSet~\cite{gemmeke2017audio} with expert musician annotations. FakeMusicCaps generates counterfeit music from SunoCaps-style captions using six synthesis methods (TTM01--TTM05 plus an unknown method). Real clips are split into train/val/test with a 7:1:2 ratio, while fake clips use TTM01--TTM03 for training, TTM04 for validation, and TTM05 plus the unknown method for testing.
\subsection{Baselines}
We evaluate the ALLM under an all-type training and evaluation protocol, including Qwen2-Audio-Chat-7B\footnote{https://huggingface.co/Qwen/Qwen2-Audio-7B-Instruct}, Qwen2.5-Omni-3B\footnote{https://huggingface.co/Qwen/Qwen2.5-Omni-3B}, and Qwen2.5-Omni-7B\footnote{https://huggingface.co/Qwen/Qwen2.5-Omni-7B}.
In the FT-GRPO stage, we select Qwen2.5-Omni-3B for experiments, balancing effectiveness and efficiency.
For SOTA comparisons, we include a prior small-model all-type ADD approach~\cite{xie2025detect} and a recent large-model approach~\cite{gu2025allm4add}.

\subsection{Implemenation Details}
All experiments are conducted on 8 NVIDIA A100 GPUs, and our fine-tuning pipeline is implemented with ms-swift~\cite{zhao2025swift}.
For both SFT and GRPO, we adopt LoRA~\cite{hu2022lora} for parameter-efficient fine-tuning, with LoRA rank \(r=64\), LoRA alpha (scaling factor) \(=16\), and LoRA dropout \(=0.05\), following the settings explored in prior work~\cite{gu2025allm4add}.
During training, the LLM and the audio encoder are both set to be trainable.
For the SFT baselines, we train on each audio type for 5 epochs.
For FT-GRPO, the Step-1 SFT cold start uses 3 epochs for speech and music, and 2 epochs for sound and singing, while Step-2 GRPO uses 2 epochs for all types. For both FT-GRPO two steps, we use a learning rate of \(1\times10^{-5}\) and a batch size per device of 16. In the Step-2 GRPO, we set the $G$ to 8, the sampling temperature to 0.9. In the total reward formulation, both $\alpha$ and $\beta$ are set to 0.1.
As for evaluation, all experiments report accuracy (ACC), computed by directly matching the predicted label in the model response with the ground-truth label.
Compared with EER, which is commonly used in prior ADD benchmarks, ACC better reflects real-world decision making.

\section{Results and Analyse}
\subsection{ALLM Performance on All-Type ADD}
ALLMs are typically trained on heterogeneous audio data and multiple tasks; however, their effectiveness on ADD task remains unclear.
We first evaluate representative ALLMs under the protocol described in Section~\ref{dataset}, and report the results in Table~\ref{tab:alltype_results}.

When trained on speech data, Qwen2.5-Omni-3B achieves the best in-domain (ID) accuracy of 99.04\% and the highest average performance of 82.45\%, indicating relatively strong generalization to other audio types.
In contrast, for the other three training types (sound, singing, and music), none of the evaluated ALLMs consistently attain high ID performance, suggesting that ADD is not reliably solved by generic all-purpose audio instruction tuning.
Notably, the sound-trained CMs exhibits the largest degradation, with the ID accuracy (Sound column) remaining only around the \(\sim\)60\% level for all models (62.60--67.55\%), highlighting the difficulty of environmental-sound deepfake detection and the limited generalizability of current ALLMs to detect sound type.

\subsection{Ablation study for FT-GRPO}
\begin{figure}[!t]
	\centering
	\subfloat{\includegraphics[width=2.5in]{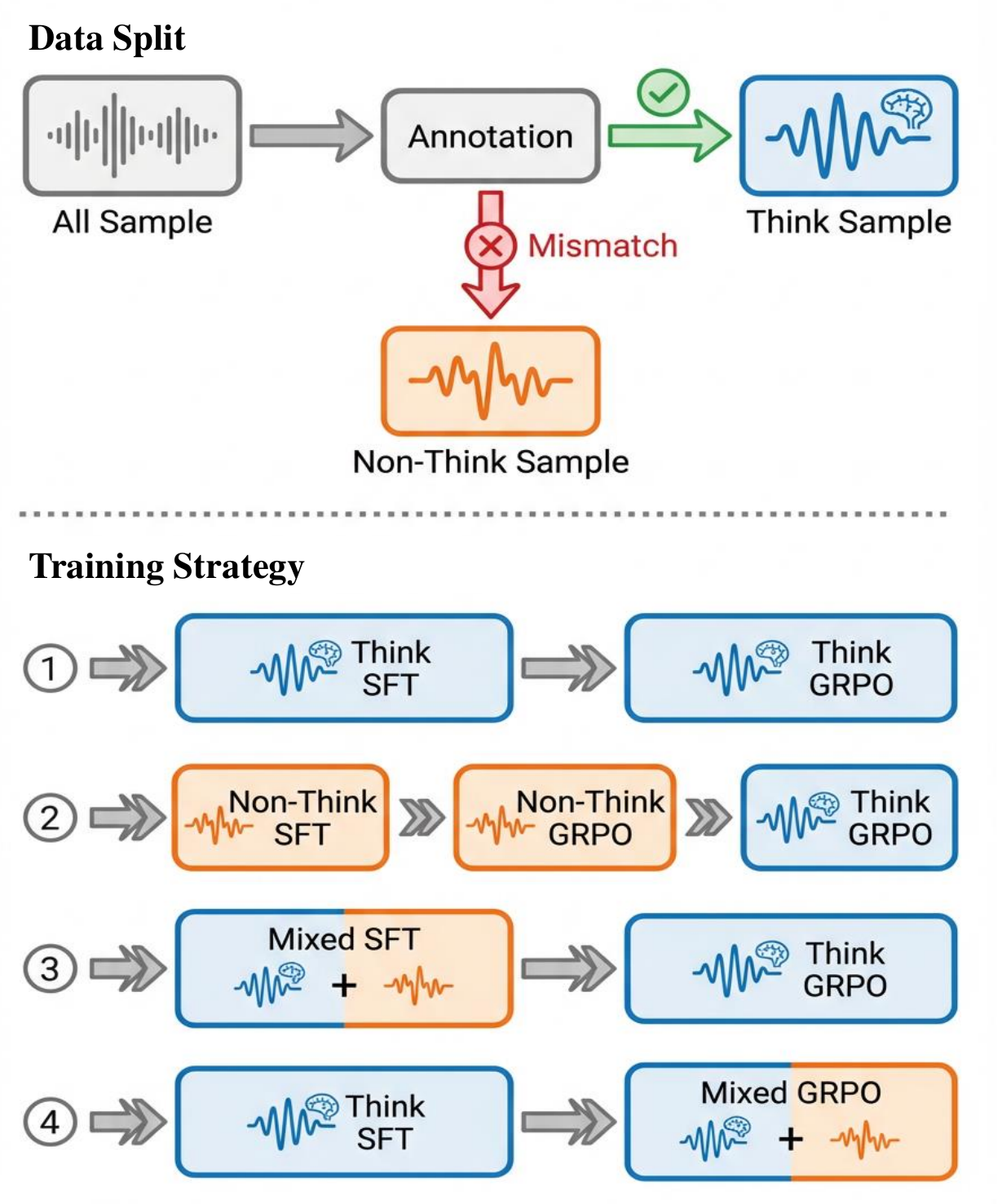}}
	\hfil
	\caption{Four different training strategies in FT-GRPO.}
	\label{fig:ablation} 
\end{figure}

\begin{table}[t]
\centering
\small
\setlength{\tabcolsep}{4pt}
\renewcommand{\arraystretch}{1.08}
\begin{tabular}{lccccc}
\toprule
\textbf{Strategy} & \textbf{Speech} & \textbf{Sound} & \textbf{Singing} & \textbf{Music} & \textbf{AVG} \\
\midrule
SFT  & 99.04 & 73.31 & 66.29 & \bf 91.16 & 82.45 \\
GRPO &96.32	&68.14	&78.21	 &88.75	 &82.86 \\
S\ding{172}   & 99.30 & 71.75 & 83.24 & 89.76 & 86.01 \\
S\ding{173}   & 98.36 & 66.94 & \bf 85.43 & 89.67 & 85.10 \\
S\ding{174}   & 99.46 & 71.68 & 83.65 & 90.01 & 86.20 \\
S\ding{175} & \bf 99.75 &\bf 75.05 & 84.26 & 90.44 &\bf 87.38 \\
\bottomrule
\end{tabular}
\caption{Ablation results on different training strategies (\%) for speech-trained Qwen2.5-Omni-3B.}
\label{tab:ablation_ftgrpo}
\end{table}

\begin{table*}[t]
\centering
\small
\setlength{\tabcolsep}{4pt}
\renewcommand{\arraystretch}{1.08}
\begin{tabular}{lccccc c}
\toprule
\textbf{Model} & \textbf{Speech} & \textbf{Sound} & \textbf{Singing} & \textbf{Music} & \textbf{Explainable} & \textbf{AVG} \\
\midrule
W2V2-AASIST \cite{tak2022automatic}      & 99.58 & 37.17 & 59.08 & 58.23 & $\times$ & 63.52 \\
WPT-W2V2-AASIST \cite{xie2025detect}   & 99.06 & 46.37 & 38.56 & 92.87 & $\times$ & 69.22 \\
ALLM4ADD \cite{gu2025allm4add}        & 99.43 & 72.00 & 80.85 & \bf 91.09 & $\times$ & 85.84 \\
FT-GRPO            & \bf 99.75 & \bf 75.05 & \bf 84.26 & 90.44 & $\checkmark$ &\bf 87.38 \\
\bottomrule
\end{tabular}
\caption{Performance comparison with the SOTA methods.}
\label{tab:sota_compare}
\vspace{-2mm}
\end{table*}
\begin{table*}[t]
\centering
\small
\setlength{\tabcolsep}{4.2pt}
\renewcommand{\arraystretch}{1.08}
\begin{tabular}{llccccc}
\toprule
\textbf{Train Type} & \textbf{Method} & \textbf{Speech} & \textbf{Sound} & \textbf{Singing} & \textbf{Music} & \textbf{AVG} \\
\midrule

Speech  & SFT     & 99.04 & 73.31 & 66.29 & 91.16 & 82.45 \\
\rowcolor{gray!10}
Speech  & FT-GRPO & \textbf{99.75}\chg{+0.99} & \textbf{75.05}\chg{+1.74} & \textbf{84.26}\chg{+17.97} & 90.44\chg{-0.72} & \textbf{87.38}\chg{+5.00} \\
\midrule

Sound   & SFT     & 86.70 & 67.55 & 77.99 & 93.58 & 81.46 \\
\rowcolor{gray!10}
Sound   & FT-GRPO & \textbf{89.68}\chg{+2.98} & \textbf{75.00}\chg{+7.45} & \textbf{85.34}\chg{+7.35} & 91.21\chg{-2.37} & \textbf{85.31}\chg{+3.85} \\
\midrule

Singing & SFT     & 39.78 & 67.13 & 83.97 & 73.39 & 66.07 \\
\rowcolor{gray!10}
Singing & FT-GRPO & \textbf{57.98}\chg{+18.20} & \textbf{75.02}\chg{+7.89} & \textbf{86.58}\chg{+2.61} & \textbf{90.58}\chg{+17.19} & \textbf{77.54}\chg{+11.47} \\
\midrule

Music   & SFT     & \textbf{83.72} & \textbf{73.40} & 23.06 & 77.34 & 64.38 \\
\rowcolor{gray!10}
Music   & FT-GRPO & 60.54\chg{-23.18} & 59.11\chg{-14.29} & \textbf{60.49}\chg{+37.43} & \textbf{83.57}\chg{+6.23} & \textbf{65.93}\chg{+1.55} \\
\midrule

Co-train & SFT     & 98.55 & 68.17 & 79.73 & 93.34 & 84.95 \\
\rowcolor{gray!10}
Co-train & FT-GRPO & \textbf{98.74}\chg{+0.19} & \textbf{77.15}\chg{+8.98} & \textbf{86.16}\chg{+6.43} & \textbf{98.33}\chg{+4.99} & \textbf{90.10}\chg{+5.15} \\
\bottomrule
\end{tabular}
\caption{Comparison of SFT and FT-GRPO across all audio type types (\%).}
\label{tab:compare_sft_ftgrpo}
\vspace{-5mm}
\end{table*}
Given the limitations of SFT-only training---including suboptimal performance and limited interpretability---we further explore RL for ADD. As this strategy has been shown to outperform vanilla SFT in the vision domain~\cite{liu2025visual}, we first apply GRPO directly to the speech-trained Qwen2.5-Omni-3B to examine its effectiveness. The results are reported in Table~\ref{tab:ablation_ftgrpo}. Compared with SFT-only training, GRPO yields a modest \(0.48\%\) improvement in average ACC; however, its ID Speech accuracy drops to \(96.32\%\). Moreover, inspection of the generated outputs shows that, despite the FT reward and a prescribed thinking pattern, the reasoning remains weakly constrained and often degenerates into irrelevant or noisy statements, likely due to reward hacking or shortcut behavior~\cite{baker2025monitoring}. This is expected because our reward mainly targets prediction correctness and format compliance, without explicitly supervising the reasoning trajectory. Inspired by the cold-start + RL paradigm in DeepSeek-R1, we therefore bootstrap FT structured CoT reasoning using cold-start demonstrations to provide a more stable initialization for subsequent GRPO, improving readability and reducing degenerate reasoning.

This raises another issue: how should we handle mismatched samples during the data annotation stage? In particular, will samples without thinking process affect performance? To answer this, we conduct four ablation training strategies to study where to place these non-think samples, as illustrated in Fig.~\ref{fig:ablation}. Specifically, S\ding{172} discards the non-think samples; S\ding{173} consists of three stages: it first performs SFT on the non-think samples, then continues SFT on the thinking annotated samples , and finally applies GRPO; S\ding{174} mixes non-think samples into the SFT stage; and S\ding{175} conducts SFT using only thinking annotated samples, while applying GRPO with the mixed data in the RL stage. The results are shown in Table~\ref{tab:ablation_ftgrpo}. We find that S\ding{175} achieves the best performance, reaching the highest ID Speech accuracy of 99.75\% and the best average accuracy of 87.38\%. Furthermore, we observe that injecting samples without thinking annotations during the cold-start stage substantially degrades token-level accuracy during SFT, indicating poorer adherence to the desired structured rationale format, which may hinder subsequent reasoning-style learning.

\subsection{Compared with SOTA Models}
After the ablation study, we compare the speech-trained FT-GRPO-Qwen2.5-Omni-3B with existing SOTA methods. As shown in Table~\ref{tab:sota_compare}, our FT-GRPO consistently outperforms both prior small-model baselines (e.g., W2V2-AASIST and WPT-W2V2-AASIST) and previous large-model approaches. In addition to improved explainability, FT-GRPO achieves the best Speech accuracy of \(99.75\%\) and the highest average accuracy of \(87.38\%\).
Notably, the reported accuracies of small models are obtained using a fixed decision threshold of 0.5, whereas ALLM4ADD is evaluated by directly matching the model responses with ground-truth labels. This also suggests a practical limitation of small models: selecting an appropriate threshold---especially under out-of-domain conditions---can be non-trivial.

\subsection{FT-GRPO for all-type ADD task}
In this section, we extend FT-GRPO from the speech-trained to the all-type ADD task. The results are reported in Table~\ref{tab:compare_sft_ftgrpo}.
Across all audio types, FT-GRPO consistently outperforms SFT-only training. Specifically, when trained on Speech, FT-GRPO improves the average accuracy by \(+5.00\%\), with particularly large gains on Singing (\(+17.97\%\)). When trained on Sound, FT-GRPO increases the average accuracy by \(+3.85\%\). When trained on Singing, FT-GRPO yields the largest overall improvement, boosting the average accuracy by \(+11.47\%\), and improving Speech and Music by \(+18.20\%\) and \(+17.19\%\), respectively.
In contrast, training on Music leads to a relatively small gain in average accuracy (\(+1.55\%\). This is mainly because the improvements on Singing (\(+37.43\%\)) and Music (\(+6.23\%\)) are partially offset by performance drops on Speech (\(-23.18\%\)) and Sound (\(-14.29\%\)), suggesting weaker cross-type generalization when trainging on Music alone.
Finally, in the co-train setting, co-trained FT-GRPO achieves the best overall performance with an average accuracy of \(90.10\%\) (a \(+5.15\%\) gain over SFT). This improvement is primarily driven by substantial gains on Sound (\(+8.98\%\)) and Music (\(+4.99\%\)), together with consistent improvements on Singing (\(+6.43\%\)) and Speech (\(+0.19\%\)).

\vspace{-1mm}
\section{Conclusion}
\vspace{-1mm}
We study all-type ADD task from the perspective of ALLMs, with the goal of achieving both strong cross-type performance and interpretable decisions. We first establish an all-type ADD benchmark and conduct a systematic evaluation of representative ALLMs. To improve explainable of ALLMs, we propose an automatic annotation-and-cleaning pipeline that generates FT CoT rationales at scale, yielding 340K high-quality instances. Building on these data, we introduce FT-GRPO, a two-stage reinforcement fine-tuning framework with an SFT cold start followed by GRPO under composite rewards for correctness, format compliance, and complete FT reasoning, while incorporating non-think samples where ALLMs tend to fail. Experiments show that FT-GRPO achieves SOTA accuracy with a 3B ALLM and further improves all-type performance under co-training.

\section*{Limitations}
Although this work demonstrates that ALLM models can achieve SOTA performance on all-type ADD task, several limitations remain to be addressed. First, our experiments cover only a limited set of ALLMs, primarily from the Qwen-Audio/Omni family, and our conclusions may not fully generalize to other architectures. Second, while our rationales are generated through a two-stage automatic annotation-and-polishing pipeline with consistency checks and cleaning, some annotations may still be suboptimal due to the limited audio understanding capabilities of current ALLMs. Third, although we evaluate on four widely used public datasets covering major audio types, the test conditions are not exhaustive, e.g., in acoustic environments, synthesis methods, partially spoofed audio, or mixed audio type scenarios. Addressing these limitations will be a focus of our future work.

\bibliography{myrefs}

@article{liu2023asvspoof,
  title={Asvspoof 2021: Towards spoofed and deepfake speech detection in the wild},
  author={Liu, Xuechen and Wang, Xin and Sahidullah, Md and Patino, Jose and Delgado, H{\'e}ctor and Kinnunen, Tomi and Todisco, Massimiliano and Yamagishi, Junichi and Evans, Nicholas and Nautsch, Andreas and others},
  journal={IEEE/ACM Transactions on Audio, Speech, and Language Processing},
  year={2023},
  publisher={IEEE}
}

@inproceedings{jung2022aasist,
  title={Aasist: Audio anti-spoofing using integrated spectro-temporal graph attention networks},
  author={Jung, Jee-weon and Heo, Hee-Soo and Tak, Hemlata and Shim, Hye-jin and Chung, Joon Son and Lee, Bong-Jin and Yu, Ha-Jin and Evans, Nicholas},
  booktitle={Proceedings of the ICASSP},
  pages={6367--6371},
  year={2022}
}

@article{tak2021end1,
  title={End-to-end spectro-temporal graph attention networks for speaker verification anti-spoofing and speech deepfake detection},
  author={Tak, H. , Jee-weon and Patino, J. and Kamble, M. and Todisco, M. and Evans, N. },
  journal={arXiv preprint arXiv:2107.12710},
  year={2021}
}

@inproceedings{tak2022automatic,
  title={Automatic Speaker Verification Spoofing and Deepfake Detection Using Wav2vec 2.0 and Data Augmentation},
  author={Tak, Hemlata and Todisco, Massimiliano and Wang, Xin and Jung, Jee-weon and Yamagishi, Junichi and Evans, Nicholas},
  booktitle={The Speaker and Language Recognition Workshop (Odyssey 2022)},
  year={2022},
  organization={ISCA}
}

@article{agostinelli2023musiclm,
  title={Musiclm: Generating music from text},
  author={Agostinelli, Andrea and Denk, Timo I and Borsos, Zal{\'a}n and Engel, Jesse and Verzetti, Mauro and Caillon, Antoine and Huang, Qingqing and Jansen, Aren and Roberts, Adam and Tagliasacchi, Marco and others},
  journal={arXiv preprint arXiv:2301.11325},
  year={2023}
}

@inproceedings{todisco19_interspeech,
  author={Massimiliano Todisco and Xin Wang and Ville Vestman and Md. Sahidullah and Héctor Delgado and Andreas Nautsch and Junichi Yamagishi and Nicholas Evans and Tomi H. Kinnunen and Kong Aik Lee},
  title={{ASVspoof 2019: Future Horizons in Spoofed and Fake Audio Detection}},
  year=2019,
  booktitle={Proc. Interspeech 2019},
  pages={1008--1012},
  doi={10.21437/Interspeech.2019-2249}
}

@article{chen2022wavlm,
  title={Wavlm: Large-scale self-supervised pre-training for full stack speech processing},
  author={Chen, Sanyuan and Wang, Chengyi and Chen, Zhengyang and Wu, Yu and Liu, Shujie and Chen, Zhuo and Li, Jinyu and Kanda, Naoyuki and Yoshioka, Takuya and Xiao, Xiong and others},
  journal={IEEE Journal of Selected Topics in Signal Processing},
  volume={16},
  number={6},
  pages={1505--1518},
  year={2022},
  publisher={IEEE}
}

@inproceedings{zhang2024audio,
  title={Audio deepfake detection with self-supervised XLS-R and SLS classifier},
  author={Zhang, Qishan and Wen, Shuangbing and Hu, Tao},
  booktitle={ACM Multimedia 2024},
  year={2024}
}

@article{phukan2024heterogeneity,
  title={Heterogeneity over Homogeneity: Investigating Multilingual Speech Pre-Trained Models for Detecting Audio Deepfake},
  author={Phukan, Orchid Chetia and Kashyap, Gautam Siddharth and Buduru, Arun Balaji and Sharma, Rajesh},
  journal={arXiv preprint arXiv:2404.00809},
  year={2024}
}

@article{comanducci2024fakemusiccaps,
  title={Fakemusiccaps: a dataset for detection and attribution of synthetic music generated via text-to-music models},
  author={Comanducci, Luca and Bestagini, Paolo and Tubaro, Stefano},
  journal={arXiv preprint arXiv:2409.10684},
  year={2024}
}

@article{li2024audio,
  title={From Audio Deepfake Detection to AI-Generated Music Detection--A Pathway and Overview},
  author={Li, Yupei and Milling, Manuel and Specia, Lucia and Schuller, Bj{\"o}rn W},
  journal={arXiv preprint arXiv:2412.00571},
  year={2024}
}

@article{li2024detecting,
  title={Detecting Machine-Generated Music with Explainability--A Challenge and Early Benchmarks},
  author={Li, Yupei and Sun, Qiyang and Li, Hanqian and Specia, Lucia and Schuller, Bj{\"o}rn W},
  journal={arXiv preprint arXiv:2412.13421},
  year={2024}
}

@inproceedings{wei2025voices,
  title={From Voices to Beats: Enhancing Music Deepfake Detection by Identifying Forgeries in Background},
  author={Wei, Zhaolin and Ye, Dengpan and Deng, Jiacheng and Lin, Yuhan},
  booktitle={ICASSP 2025-2025 IEEE International Conference on Acoustics, Speech and Signal Processing (ICASSP)},
  pages={1--5},
  year={2025},
  organization={IEEE}
}

@inproceedings{li2024mert,
  title={MERT: Acoustic Music Understanding Model with Large-Scale Self-supervised Training},
  author={Li, Yizhi and Yuan, Ruibin and Zhang, Ge and Ma, Yinghao and Chen, Xingran and Yin, Hanzhi and Xiao, Chenghao and Lin, Chenghua and Ragni, Anton and Benetos, Emmanouil and others},
  booktitle={ICLR},
  year={2024}
}

@inproceedings{wang2024asvspoof5,
  title = {{ASVspoof 5}: {Crowdsourced} Speech Data, Deepfakes, and Adversarial Attacks at Scale},
  booktitle = {ASVspoof Workshop 2024 (accepted)},
  author = {Wang, Xin and Delgado, H{\'e}ctor and Tak, Hemlata and Jung, Jee-weon and Shim, Hye-jin and Todisco, Massimiliano and Kukanov, Ivan and Liu, Xuechen and Sahidullah, Md and Kinnunen, Tomi and Evans, Nicholas and Lee, Kong Aik and Yamagishi, Junichi},
  year = {2024}
}

@inproceedings{zhang2024svdd,
  title={Svdd 2024: The inaugural singing voice deepfake detection challenge},
  author={Zhang, You and Zang, Yongyi and Shi, Jiatong and Yamamoto, Ryuichi and Toda, Tomoki and Duan, Zhiyao},
  booktitle={2024 IEEE Spoken Language Technology Workshop (SLT)},
  pages={782--787},
  year={2024},
  organization={IEEE}
}

@inproceedings{xie2024fsd,
  title={Fsd: An initial chinese dataset for fake song detection},
  author={Xie, Yuankun and Zhou, Jingjing and Lu, Xiaolin and Jiang, Zhenghao and Yang, Yuxin and Cheng, Haonan and Ye, Long},
  booktitle={ICASSP 2024-2024 IEEE International Conference on Acoustics, Speech and Signal Processing (ICASSP)},
  pages={4605--4609},
  year={2024},
  organization={IEEE}
}

@article{zang2024ctrsvdd,
  title={Ctrsvdd: A benchmark dataset and baseline analysis for controlled singing voice deepfake detection},
  author={Zang, Yongyi and Shi, Jiatong and Zhang, You and Yamamoto, Ryuichi and Han, Jionghao and Tang, Yuxun and Xu, Shengyuan and Zhao, Wenxiao and Guo, Jing and Toda, Tomoki and others},
  journal={arXiv preprint arXiv:2406.02438},
  year={2024}
}

@inproceedings{xie2024fakesound,
  title={FakeSound: Deepfake General Audio Detection},
  author={Xie, Zeyu and Li, Baihan and Xu, Xuenan and Liang, Zheng and Yu, Kai and Wu, Mengyue},
  booktitle={Proc. Interspeech 2024},
  pages={112--116},
  year={2024}
}

@inproceedings{gohari2025audio,
  title={Audio Features Investigation for Singing Voice Deepfake Detection},
  author={Gohari, Mahyar and Salvi, Davide and Bestagini, Paolo and Adami, Nicola},
  booktitle={ICASSP 2025-2025 IEEE International Conference on Acoustics, Speech and Signal Processing (ICASSP)},
  pages={1--5},
  year={2025},
  organization={IEEE}
}

@inproceedings{wang2025mixture,
  title={Mixture of experts fusion for fake audio detection using frozen wav2vec 2.0},
  author={Wang, Zhiyong and Fu, Ruibo and Wen, Zhengqi and Tao, Jianhua and Wang, Xiaopeng and Xie, Yuankun and Qi, Xin and Shi, Shuchen and Lu, Yi and Liu, Yukun and others},
  booktitle={ICASSP 2025-2025 IEEE International Conference on Acoustics, Speech and Signal Processing (ICASSP)},
  pages={1--5},
  year={2025},
  organization={IEEE}
}

@article{kheir2025comprehensive,
  title={Comprehensive Layer-wise Analysis of SSL Models for Audio Deepfake Detection},
  author={Kheir, Yassine El and Samih, Youness and Maharjan, Suraj and Polzehl, Tim and M{\"o}ller, Sebastian},
  journal={arXiv preprint arXiv:2502.03559},
  year={2025}
}

@inproceedings{pan2024attentive,
  title={Attentive Merging of Hidden Embeddings from Pre-trained Speech Model for Anti-spoofing Detection},
  author={Pan, Zihan and Liu, Tianchi and Sailor, Hardik B and Wang, Qiongqiong},
  booktitle={Proc. Interspeech 2024},
  pages={2090--2094},
  year={2024}
}

@INPROCEEDINGS{10832226,
  author={Guragain, Anmol and Liu, Tianchi and Pan, Zihan and Sailor, Hardik B. and Wang, Qiongqiong},
  booktitle={2024 IEEE Spoken Language Technology Workshop (SLT)}, 
  title={Speech Foundation Model Ensembles for the Controlled Singing Voice Deepfake Detection (CTRSVDD) Challenge 2024}, 
  year={2024},
  volume={},
  number={},
  pages={774-781},
  doi={10.1109/SLT61566.2024.10832226}}

@inproceedings{
zhang2025i,
title={I Can Hear You: Selective Robust Training for Deepfake Audio Detection},
author={Zirui Zhang and Wei Hao and Aroon Sankoh and William Lin and Emanuel Mendiola-Ortiz and Junfeng Yang and Chengzhi Mao},
booktitle={The Thirteenth International Conference on Learning Representations},
year={2025},
url={https://openreview.net/forum?id=2GcR9bO620}
}

@inproceedings{kawa2023defense,
  title={Defense Against Adversarial Attacks on Audio DeepFake Detection},
  author={Kawa, Piotr and Plata, Marcin and Syga, Piotr},
  booktitle={Proc. Interspeech 2023},
  pages={5276--5280},
  year={2023}
}

@inproceedings{zhang2024xwsb,
  title={XWSB: A Blend System Utilizing XLS-R and Wavlm With SLS Classifier Detection System for SVDD 2024 Challenge},
  author={Zhang, Qishan and Wen, Shuangbing and Yan, Fangke and Hu, Tao and Li, Jun},
  booktitle={2024 IEEE Spoken Language Technology Workshop (SLT)},
  pages={788--794},
  year={2024},
  organization={IEEE}
}

@inproceedings{chen2024singing,
  title={Singing Voice Graph Modeling for SingFake Detection},
  author={Chen, Xuanjun and Wu, Haibin and Jang, Roger and Lee, Hung-yi},
  booktitle={Proc. Interspeech 2024},
  pages={4843--4847},
  year={2024}
}

@inproceedings{gemmeke2017audio,
  title={Audio set: An ontology and human-labeled dataset for audio events},
  author={Gemmeke, Jort F and Ellis, Daniel PW and Freedman, Dylan and Jansen, Aren and Lawrence, Wade and Moore, R Channing and Plakal, Manoj and Ritter, Marvin},
  booktitle={2017 IEEE international conference on acoustics, speech and signal processing (ICASSP)},
  pages={776--780},
  year={2017},
  organization={IEEE}
}

@inproceedings{yin25_interspeech,
  title     = {{EnvSDD: Benchmarking Environmental Sound Deepfake Detection}},
  author    = {Han Yin and Yang Xiao and Rohan Kumar Das and Jisheng Bai and Haohe Liu and Wenwu Wang and Mark D Plumbley},
  year      = {2025},
  booktitle = {{Interspeech 2025}},
  pages     = {201--205},
  doi       = {10.21437/Interspeech.2025-1143},
  issn      = {2958-1796},
}

@article{xie2025detect,
  title={Detect All-Type Deepfake Audio: Wavelet Prompt Tuning for Enhanced Auditory Perception},
  author={Xie, Yuankun and Fu, Ruibo and Wang, Zhiyong and Wang, Xiaopeng and Cao, Songjun and Ma, Long and Cheng, Haonan and Ye, Long},
  journal={arXiv preprint arXiv:2504.06753},
  year={2025}
}

@inproceedings{tang2024extending,
  title={Extending large language models for speech and audio captioning},
  author={Tang, Changli and Yu, Wenyi and Sun, Guangzhi and Chen, Xianzhao and Tan, Tian and Li, Wei and Lu, Lu and Ma, Zejun and Zhang, Chao},
  booktitle={ICASSP 2024-2024 IEEE International Conference on Acoustics, Speech and Signal Processing (ICASSP)},
  pages={11236--11240},
  year={2024},
  organization={IEEE}
}

@inproceedings{zhifei-etal-2025-audio,
    title = "Audio-Reasoner: Improving Reasoning Capability in Large Audio Language Models",
    author = "Zhifei, Xie  and
      Lin, Mingbao  and
      Liu, Zihang  and
      Wu, Pengcheng  and
      Yan, Shuicheng  and
      Miao, Chunyan",
    editor = "Christodoulopoulos, Christos  and
      Chakraborty, Tanmoy  and
      Rose, Carolyn  and
      Peng, Violet",
    booktitle = "Proceedings of the 2025 Conference on Empirical Methods in Natural Language Processing",
    month = nov,
    year = "2025",
    address = "Suzhou, China",
    publisher = "Association for Computational Linguistics",
    url = "https://aclanthology.org/2025.emnlp-main.1216/",
    doi = "10.18653/v1/2025.emnlp-main.1216",
    pages = "23840--23862",
    ISBN = "979-8-89176-332-6"
}

@article{ma2025mmar,
  title={MMAR: A Challenging Benchmark for Deep Reasoning in Speech, Audio, Music, and Their Mix},
  author={Ma, Ziyang and Ma, Yinghao and Zhu, Yanqiao and Yang, Chen and Chao, Yi-Wen and Xu, Ruiyang and Chen, Wenxi and Chen, Yuanzhe and Chen, Zhuo and Cong, Jian and others},
  journal={arXiv preprint arXiv:2505.13032},
  year={2025}
}

@article{wang2025speechllm,
  title={SpeechLLM-as-Judges: Towards General and Interpretable Speech Quality Evaluation},
  author={Wang, Hui and Zhao, Jinghua and Yang, Yifan and Liu, Shujie and Chen, Junyang and Zhang, Yanzhe and Zhao, Shiwan and Li, Jinyu and Zhou, Jiaming and Sun, Haoqin and others},
  journal={arXiv preprint arXiv:2510.14664},
  year={2025}
}

@article{li2025audiotrust,
  title={Audiotrust: Benchmarking the multifaceted trustworthiness of audio large language models},
  author={Li, Kai and Shen, Can and Liu, Yile and Han, Jirui and Zheng, Kelong and Zou, Xuechao and Wang, Zhe and Zhang, Shun and Du, Xingjian and Luo, Hanjun and others},
  journal={arXiv preprint arXiv:2505.16211},
  year={2025}
}

@article{yang2025towards,
  title={Towards holistic evaluation of large audio-language models: A comprehensive survey},
  author={Yang, Chih-Kai and Ho, Neo S and Lee, Hung-yi},
  journal={arXiv preprint arXiv:2505.15957},
  year={2025}
}

@article{zhang2024speechgpt,
  title={Speechgpt-gen: Scaling chain-of-information speech generation},
  author={Zhang, Dong and Zhang, Xin and Zhan, Jun and Li, Shimin and Zhou, Yaqian and Qiu, Xipeng},
  journal={arXiv preprint arXiv:2401.13527},
  year={2024}
}

@article{yan2025ming,
  title={Ming-UniAudio: Speech LLM for Joint Understanding, Generation and Editing with Unified Representation},
  author={Yan, Canxiang and Jin, Chunxiang and Huang, Dawei and Yu, Haibing and Peng, Han and Zhan, Hui and Gao, Jie and Peng, Jing and Chen, Jingdong and Zhou, Jun and others},
  journal={arXiv preprint arXiv:2511.05516},
  year={2025}
}

@inproceedings{fathullah2024prompting,
  title={Prompting large language models with speech recognition abilities},
  author={Fathullah, Yassir and Wu, Chunyang and Lakomkin, Egor and Jia, Junteng and Shangguan, Yuan and Li, Ke and Guo, Jinxi and Xiong, Wenhan and Mahadeokar, Jay and Kalinli, Ozlem and others},
  booktitle={ICASSP 2024-2024 IEEE International Conference on Acoustics, Speech and Signal Processing (ICASSP)},
  pages={13351--13355},
  year={2024},
  organization={IEEE}
}

@article{wang2025qualispeech,
  title={Qualispeech: A speech quality assessment dataset with natural language reasoning and descriptions},
  author={Wang, Siyin and Yu, Wenyi and Chen, Xianzhao and Tian, Xiaohai and Zhang, Jun and Lu, Lu and Tsao, Yu and Yamagishi, Junichi and Wang, Yuxuan and Zhang, Chao},
  journal={arXiv preprint arXiv:2503.20290},
  year={2025}
}

@inproceedings{gu2025allm4add,
  title={ALLM4ADD: Unlocking the Capabilities of Audio Large Language Models for Audio Deepfake Detection},
  author={Gu, Hao and Yi, Jiangyan and Wang, Chenglong and Tao, Jianhua and Lian, Zheng and He, Jiayi and Ren, Yong and Chen, Yujie and Wen, Zhengqi},
  booktitle={Proceedings of the 33rd ACM International Conference on Multimedia},
  pages={11736--11745},
  year={2025}
}

@article{guo2025deepseek,
  title={Deepseek-r1: Incentivizing reasoning capability in llms via reinforcement learning},
  author={Guo, Daya and Yang, Dejian and Zhang, Haowei and Song, Junxiao and Zhang, Ruoyu and Xu, Runxin and Zhu, Qihao and Ma, Shirong and Wang, Peiyi and Bi, Xiao and others},
  journal={arXiv preprint arXiv:2501.12948},
  year={2025}
}

@article{zhou2025r1,
  title={R1-Zero's" Aha Moment" in Visual Reasoning on a 2B Non-SFT Model},
  author={Zhou, Hengguang and Li, Xirui and Wang, Ruochen and Cheng, Minhao and Zhou, Tianyi and Hsieh, Cho-Jui},
  journal={arXiv preprint arXiv:2503.05132},
  year={2025}
}

@inproceedings{zhao2025swift,
  title={Swift: a scalable lightweight infrastructure for fine-tuning},
  author={Zhao, Yuze and Huang, Jintao and Hu, Jinghan and Wang, Xingjun and Mao, Yunlin and Zhang, Daoze and Jiang, Zeyinzi and Wu, Zhikai and Ai, Baole and Wang, Ang and others},
  booktitle={Proceedings of the AAAI Conference on Artificial Intelligence},
  volume={39},
  number={28},
  pages={29733--29735},
  year={2025}
}

@inproceedings{
hu2022lora,
title={Lo{RA}: Low-Rank Adaptation of Large Language Models},
author={Edward J Hu and yelong shen and Phillip Wallis and Zeyuan Allen-Zhu and Yuanzhi Li and Shean Wang and Lu Wang and Weizhu Chen},
booktitle={International Conference on Learning Representations},
year={2022},
url={https://openreview.net/forum?id=nZeVKeeFYf9}
}

@article{liu2025visual,
  title={Visual-rft: Visual reinforcement fine-tuning},
  author={Liu, Ziyu and Sun, Zeyi and Zang, Yuhang and Dong, Xiaoyi and Cao, Yuhang and Duan, Haodong and Lin, Dahua and Wang, Jiaqi},
  journal={arXiv preprint arXiv:2503.01785},
  year={2025}
}

@article{baker2025monitoring,
  title={Monitoring reasoning models for misbehavior and the risks of promoting obfuscation},
  author={Baker, Bowen and Huizinga, Joost and Gao, Leo and Dou, Zehao and Guan, Melody Y and Madry, Aleksander and Zaremba, Wojciech and Pachocki, Jakub and Farhi, David},
  journal={arXiv preprint arXiv:2503.11926},
  year={2025}
}

@article{chen2025advancing,
  title={Advancing Multimodal Reasoning: From Optimized Cold Start to Staged Reinforcement Learning},
  author={Chen, Shuang and Guo, Yue and Su, Zhaochen and Li, Yafu and Wu, Yulun and Chen, Jiacheng and Chen, Jiayu and Wang, Weijie and Qu, Xiaoye and Cheng, Yu},
  journal={arXiv preprint arXiv:2506.04207},
  year={2025}
}

@article{wei2025advancing,
  title={Advancing Multimodal Reasoning via Reinforcement Learning with Cold Start},
  author={Wei, Lai and Li, Yuting and Zheng, Kaipeng and Wang, Chen and Wang, Yue and Kong, Linghe and Sun, Lichao and Huang, Weiran},
  journal={arXiv preprint arXiv:2505.22334},
  year={2025}
}

@article{guo2025envsslam,
  title={EnvSSLAM-FFN: Lightweight Layer-Fused System for ESDD 2026 Challenge},
  author={Guo, Xiaoxuan and Huang, Hengyan and Zhou, Jiayi and Sun, Renhe and Liu, Jian and Cheng, Haonan and Ye, Long and Zhang, Qin},
  journal={arXiv preprint arXiv:2512.20369},
  year={2025}
}
\clearpage
\appendix

\definecolor{PromptTitleBg}{RGB}{248,240,210} 
\definecolor{PromptFrame}{RGB}{0,0,0}         
\definecolor{PromptBlue}{RGB}{0,80,200}       
\definecolor{PromptRed}{RGB}{200,0,0}         

\tcbset{
  promptstyle/.style={
    enhanced,
    colback=white,
    colframe=PromptFrame,
    boxrule=0.9pt,
    arc=3mm,
    left=3mm,right=3mm,top=2mm,bottom=2mm,
    fonttitle=\bfseries,          
    colbacktitle=PromptTitleBg,
    coltitle=black,
    boxed title style={
      boxrule=0.9pt,
      colframe=PromptFrame,
      colback=PromptTitleBg,
      arc=3mm
    },
    attach boxed title to top left={xshift=0mm,yshift=-1.2mm},
  }
}

\section{Details of Data Construction}
\label{sec:appendix_data}
\begin{table*}[t]
\centering
\setlength{\tabcolsep}{5pt}
\renewcommand{\arraystretch}{1.1}
\begin{tabular}{lccccc}
\toprule
\multirow{2}{*}{\textbf{Type}} &
\multicolumn{2}{c}{\textbf{Train}} &
\multicolumn{2}{c}{\textbf{Dev}} &
\multirow{2}{*}{\textbf{Non-think rate (\%)}} \\
\cmidrule{2-3}\cmidrule{4-5}
& \textbf{Think} & \textbf{Non-Think} & \textbf{Think} & \textbf{Non-Think} & \\
\midrule
Speech  & 23,426  & 1,954  & 23,049 & 1,795 & 8.06  \\
Sound   & 121,544 & 17,511 & 34,817 & 4,893 & 14.44 \\
Singing & 77,247  & 7,157  & 40,256 & 3,369 & 8.06  \\
Music   & 15,958  & 4,903  & 4,636  & 1,422 & 37.47 \\
\bottomrule
\end{tabular}
\caption{Statistics of think and non-think samples in the training and development splits.}
\label{tab:think-hard-stats}
\end{table*}

\begin{table}[t]
  \centering
  \begin{tabular}{lccc}
    \toprule
    $lr$ & $r$ & $\alpha$ & ACC (\%) \\
    \midrule
    $5\times10^{-5}$ & 8  & 32 & 91.70 \\
    $1\times10^{-5}$ & 8  & 32 & 97.04 \\
    $5\times10^{-6}$ & 8  & 32 & 98.70 \\
    $5\times10^{-5}$ & 64 & 16 & 93.48 \\
    $1\times10^{-5}$ & 64 & 16 & \bf 99.04 \\
    $5\times10^{-6}$ & 64 & 16 & 98.71 \\
    \bottomrule
  \end{tabular}
  \caption{Hyperparameter study of learning rate ($lr$), lora rank $r$, lora alpha $\alpha$. In this table, the Audio encoder, Aligner, and LLM are all fine-tuned.}
  \label{tab:lora_lr}
\end{table}

As shown in Figures~\ref{fig:prompt}, we illustrate the prompts used in data construction. Specifically, we apply the proposed annotation pipeline to the training and development splits of 19LA, ESDD, CtrSVDD, and FakeMusicCaps. After annotation, each audio type is partitioned into a \emph{think} subset and a \emph{non-think} subset, as summarized in Table~\ref{tab:think-hard-stats}. 

The resulting non-think rate also serves as a proxy for how reliably Qwen-Omni-Captioner-30B can produce coherent FT rationales across audio types: it remains low (around 8\%) for speech and singing voice, which are dominated by human vocal content, but rises markedly for environmental sounds and music, reaching 14.44\% and 37.47\%, respectively. More broadly, this gap highlights a capability imbalance in current ALLMs: they appear stronger at semantic understanding of vocal-centric content, yet less reliable at capturing fine-grained acoustic cues and diverse non-vocal sound patterns that are critical for authenticity judgment.

\section{Details of Training Hyperparameters}
This section summarizes the training hyperparameters used throughout our experiments.
We first study the learning rate and LoRA configurations, then compare freezing vs.\ fine-tuning different modules. Finally, we report the training prompts and the full hyperparameter settings for the two-stage FT-GRPO pipeline, including the training details for cross-type training and co-training.

\subsection{Learning Rate and LoRA Configuration}
We first determine the optimization hyperparameters on Qwen2.5-Omni-3B under a fully fine-tuned setting, i.e., the audio encoder, aligner, and LLM are all trainable. Table~\ref{tab:lora_lr} reports the ID accuracy under commonly used learning rates and two representative LoRA configurations: $(r{=}8,\alpha{=}32)$, which is the default in \texttt{ms-swift} and many LLM fine-tuning frameworks, and $(r{=}64,\alpha{=}16)$, which is reported as a strong setting by Gu et al.~\cite{gu2025allm4add}. Overall, a larger LoRA rank tends to yield higher accuracy for audio deepfake detection, suggesting that higher adapter capacity is beneficial for modeling audio-specific artifacts. Meanwhile, the learning rate remains critical: overly large learning rates degrade performance, while a moderate value ($1\times10^{-5}$) achieves the best results in our study. Unless otherwise specified, we use LoRA $(r{=}64,\alpha{=}16)$ with a learning rate of $1\times10^{-5}$ in the following ablations.

\begin{figure}[t]
  \centering
  \includegraphics[width=3in]{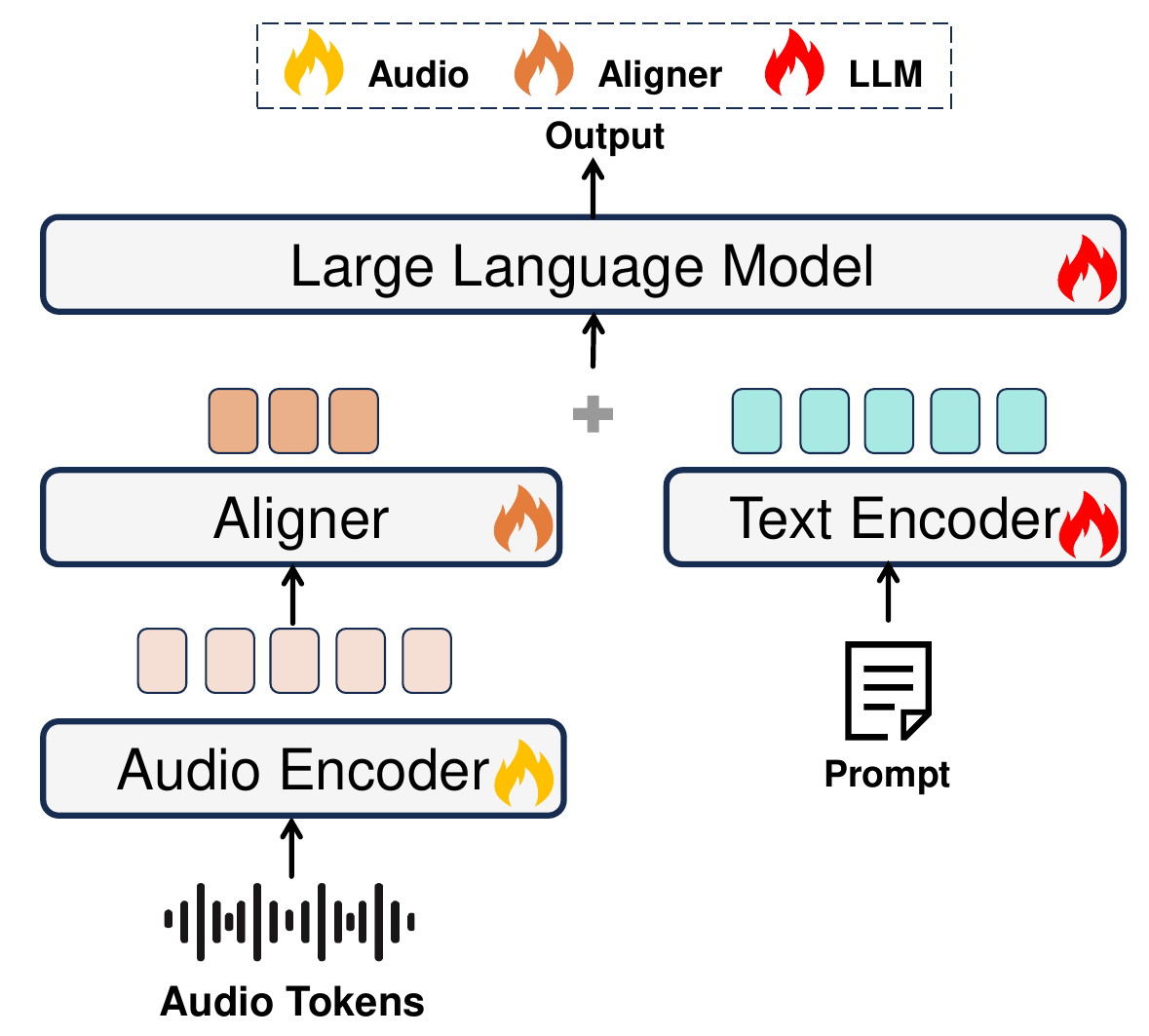}
  \caption{Three fine-tunable modules in an ALLM.}
  \label{fig:ablation}
\end{figure}

\begin{figure}[t]
  \centering
  \includegraphics[width=3in]{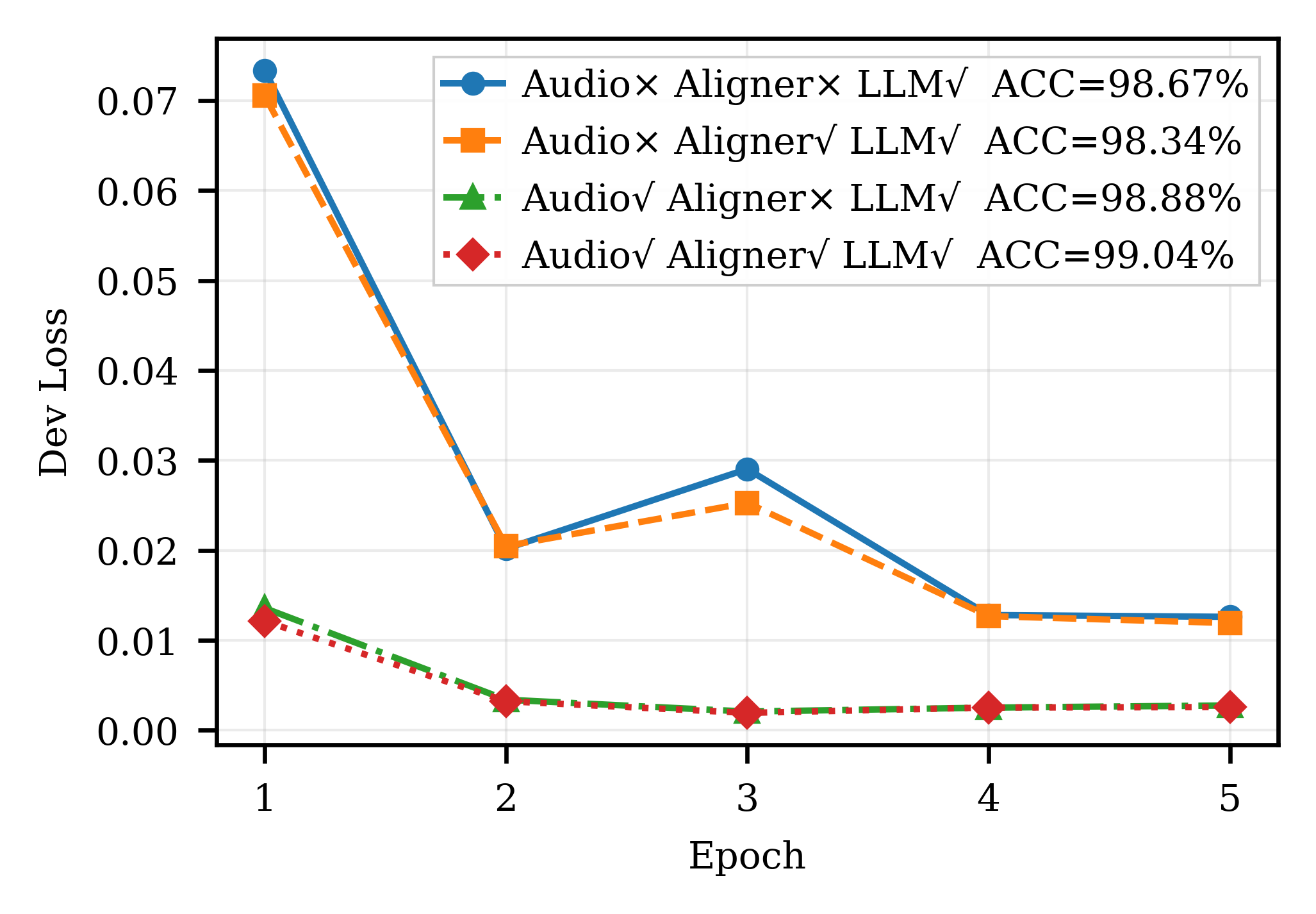}
  \caption{Four training modes for speech-trained Qwen2.5-Omni-3B. $\checkmark$ denotes trainable and $\times$ denotes frozen. ACC is measured on the 19LA evaluation set.}
  \label{fig:devloss}
\end{figure}
\subsection{Freezing vs. Fine-tuning}
As described in the main text, we adopt LoRA with an all-layer trainable setting. In practice, several module-level switches remain configurable. Figure~\ref{fig:ablation} illustrates a canonical ALLM architecture. We use \texttt{ms-swift} as our fine-tuning framework and take Qwen2.5-Omni as an example. The framework provides three freezing switches: \texttt{freeze\_vit}, \texttt{freeze\_aligner}, and \texttt{freeze\_llm}. Specifically, \texttt{freeze\_vit} controls whether the audio/vision encoders are updated; \texttt{freeze\_aligner} controls whether the modality-to-token projection (i.e., the alignment module that maps the last-layer encoder features to input tokens) is updated; and \texttt{freeze\_llm} controls whether the LLM backbone is updated.

Using the selected hyperparameters (LoRA $(r{=}64,\alpha{=}16)$ and learning rate $1\times10^{-5}$), we compare four training variants by monitoring dev loss under the 19LA train/test protocol, as shown in Figure~\ref{fig:devloss}. We observe two clearly separated loss regimes depending on whether the audio encoder is frozen or fine-tuned. Both the dev loss and the evaluation accuracy indicate that fine-tuning the audio encoder substantially outperforms freezing it. Moreover, enabling the aligner further improves performance compared to keeping it frozen. The best results are achieved when all three components are fine-tuned.

\begin{table}[t]
\centering
\small
\setlength{\tabcolsep}{6pt}
\begin{tabular}{l l}
\hline
\textbf{Stage-1 (SFT cold start)} & \textbf{Setting} \\
\hline
Base model & Qwen2.5-Omni-3B \\
Precision & bfloat16 \\
PEFT & LoRA \\
LoRA rank / alpha & 64 / 16 \\
Max sequence length & 768 \\
Epochs & 3 \\
Learning rate & $1\times 10^{-5}$ \\
Weight decay & 0.1 \\
Warmup ratio & 0.01 \\
Per-device train batch size & 16 \\
Gradient accumulation & 1 \\
DataLoader workers & 8 \\
Distributed training & DeepSpeed ZeRO-2 \\
\hline
\end{tabular}
\caption{Hyperparameters for FT-GRPO step 1.}
\label{tab:speech_sft_hparams}
\end{table}

\begin{table}[t]
\centering
\small
\setlength{\tabcolsep}{6pt}
\begin{tabular}{l l}
\hline
\textbf{Stage-2 (GRPO)} & \textbf{Setting} \\
\hline
RL method & GRPO (\texttt{rlhf\_type=grpo}) \\
Init checkpoint & SFT checkpoint\\
Precision & bfloat16 \\
PEFT & LoRA \\
LoRA rank / alpha & 64 / 16 \\
Max prompt length & 768 \\
Max completion length & 768 \\
Epochs & 2 \\
Learning rate & $1\times 10^{-5}$ \\
Weight decay & 0.1 \\
Warmup ratio & 0.05 \\
Per-device train batch size & 32 \\
Gradient accumulation & 1 \\
Group size $G$ & 8 (\texttt{num\_generations}) \\
Rewards & Accuracy + Format + FT \\
Reward weights & $[1,\ 0.1,\ 0.1]$ \\
DataLoader workers & 8 \\
Distributed training & DeepSpeed ZeRO-2 \\
\hline
\end{tabular}
\caption{Hyperparameters for FT-GRPO step 2.}
\vspace{-5mm}
\label{tab:speech_grpo_hparams}
\end{table}

\subsection{Trainging Prompt}
Similar to the data construction stage, we use FT-guided prompts in all experiments to steer the ALLM to reason from both frequency- and time-domain cues. For the real/fake SFT experiments, we slightly simplify the prompt since no strict output format is required, as shown in Figure~\ref{fig:prompt}. For FT-GRPO, both the SFT stage and the GRPO stage use the same system prompt as in the construction stage; the only change is that the user prompt is replaced with: \emph{“Could you verify whether this audio is real or fake?”}

\subsection{FT-GRPO Hyperparameters}
We list the hyperparameters for the speech-trained FT-GRPO in Step 1 (SFT cold start) and Step 2 (GRPO) in Table~\ref{tab:speech_sft_hparams} and Table~\ref{tab:speech_grpo_hparams}. For the other audio types, all hyperparameters are kept the same except for the number of epochs: in Step 1, we use 3 epochs for speech and music and 2 epochs for sound and singing, while in Step 2 we use 2 GRPO epochs for all four types. For the co-training setting, due to the much larger data scale, we use 2 epochs for the cold start and 1 epoch for GRPO.

\begin{figure*}[t]
\centering
\small
\begin{tcolorbox}[
  promptstyle,
  title=Prompt Details of data construction step 1: Raw Audio Caption,
  width=\textwidth
]
\systemtag\\
You are an expert in audio deepfake detection task.\\
The given audio is a [\textit{speech/environmental sound/singing voice/music}] clip.\\
Fake audio is entirely generated by a deepfake synthesis method, not created through manual splicing or digital editing. \\
You can detect fake audio artifacts by analyzing the audio spectrogram in two dimensions: frequency domain and time domain.

As you reason, highlight where artifacts occur to help differentiate genuine recordings from deepfakes. Give a concise thinking process by summarizing only the most important points.

Within <think>, structure your reasoning process as follows:
\begin{itemize}
  \item Start with <Frequency Domain> describe spectral characteristics and any unnatural artifacts within the valid frequency range, with close attention to the high frequencies.
  \item Follow with <Time Domain> describe temporal coherence, and any unnatural artifacts, particularly those appearing in specific time intervals, with special focus on silent or low-energy segments.
\end{itemize}

The reasoning process and answer are enclosed within <think> </think> and <answer> </answer> tags, respectively, i.e.,
<think> reasoning process here </think><answer> answer here </answer>.\\\\
\textbf{Format example} (must follow this structure exactly):\\
<think>\\
<Frequency Domain>[your analysis here]\\
<Time Domain>[your analysis here]\\
</think>\\
<answer>[\textit{real/fake}]</answer>\\\\
\prompttag

<audio>This is a [\textit{real/fake}] audio. Please provide your
thinking process within <think> </think> tags and your final answer
(real/fake) within <answer> </answer> tags.
\end{tcolorbox}
\vspace{4pt}

\begin{tcolorbox}[
  promptstyle,
  title=Prompt details of data construction step 2: Caption Polish,
  width=\textwidth
]

\systemtag\\
Your task is polish the reason content inside <think> in order to better highlight the key evidence and determine whether the real/fake description in the <think> section is consistent with the real/fake label in the <answer> section.\\\\
\textbf{Important Instructions:}\\
1.  Polish Reason: Keep only the descriptions relevant to authenticity judgment from the two dimensions. Remove all think descriptions related to editing, as there was no editing involved; \\
2.  Verify Consistency with label: If the user message indicates "this is a real/fake audio" but the description or final answer contradicts it, output "mismatch".\\
3.  Verify Internal Consistency: If the evidence in the description in `<think>` contradicts the final answer, output "mismatch".\\\\
\textbf{Format example} (must follow this structure exactly):\\
<think>\\
<Frequency Domain>[your analysis here]\\
<Time Domain>[your analysis here]\\
</think>\\
<answer>[\textit{real/fake}]</answer>\\\\
\prompttag\\
This is an authenticity description for an audio sample. Please strictly follow the important instructions to polish it.
\end{tcolorbox}

\vspace{4pt}

\begin{tcolorbox}[
  promptstyle,
  title=Training prompt in SFT step,
  width=\textwidth
]

\systemtag\\
As an AI expert in audio deepfake detection, you can identify real and deepfake audio.\\
You can detect artifacts by analyzing the spectrogram along two key dimensions—frequency domain and time domain to effectively distinguish between genuine and deepfake audio. \\
\prompttag\\
<audio>Could you verify whether this audio is real or fake?
\end{tcolorbox}
\caption{Prompt details of data construction and SFT step.}
\label{fig:prompt}
\end{figure*}

\definecolor{PromptGreen}{RGB}{0,120,60}  
\definecolor{PromptGray}{RGB}{200,0,0}    
\definecolor{RealTitleBg}{RGB}{225,245,232}
\definecolor{FakeTitleBg}{RGB}{255,228,232}
\tcbset{
  realexample/.style={
    promptstyle,
    colframe=PromptGreen,
    coltext=black,
    colback=black!1,
    coltitle=black,
    colbacktitle=RealTitleBg,
    fonttitle=\bfseries,
    boxrule=0.9pt,
    boxed title style={
      enhanced,
      boxrule=0.9pt,
      colframe=PromptGreen,
      colback=RealTitleBg,
      arc=3mm,
    },
  },
  fakeexample/.style={
    promptstyle,
    colframe=PromptGray,
    coltext=black,
    colback=black!1,
    coltitle=black,
    colbacktitle=FakeTitleBg,
    fonttitle=\bfseries,
    boxrule=0.9pt,
    boxed title style={
      enhanced,
      boxrule=0.5pt,
      colframe=PromptGray,
      colback=FakeTitleBg,
      arc=3mm,
    },
  },
}

\clearpage
\begin{figure*}[t]
\centering
\small

\begin{tcolorbox}[
  realexample,
  title={Real Example (Speech)},
  width=\textwidth
]
\textbf{File:} \texttt{LA\_T\_1078395.flac}\\[1mm]
\texttt{\textless think\textgreater}\\
\textbf{[Frequency Domain]} The audio exhibits a natural, full-frequency spectral profile with clear presence across low, mid, and high frequencies, consistent with genuine recordings. No high-frequency artifacts or distortions typically associated with deepfake synthesis are observed.\\
\textbf{[Time Domain]} The vocal rhythm, intonation, and characteristics remain stable and realistic throughout the audio. No abrupt anomalies, unnatural pauses, or inconsistencies in speech patterns that would suggest deepfake generation are detected.\\
\texttt{\textless/think\textgreater}\\
\texttt{\textless answer\textgreater}real\texttt{\textless/answer\textgreater}
\end{tcolorbox}

\vspace{2mm}

\begin{tcolorbox}[
  fakeexample,
  title={Fake Example (Speech)},
  width=\textwidth
]
\textbf{File:} \texttt{LA\_T\_7794385.flac}\\[1mm]
\texttt{\textless think\textgreater}\\
\textbf{[Frequency Domain]} The audio contains unnatural, high-frequency artifacts in the upper mid and high frequencies, which are inconsistent with natural human speech. These artifacts exhibit a synthetic spectral envelope lacking the smoothness and natural variation seen in genuine voice recordings, indicative of deepfake synthesis.\\
\textbf{[Time Domain]} The audio displays timing irregularities, such as abrupt pauses after ``What'' and unnatural cadence in phrases like ``in the game.'' These anomalies reflect algorithmic generation patterns rather than organic speech rhythms, further supporting synthetic origin.\\
\texttt{\textless/think\textgreater}\\
\texttt{\textless answer\textgreater}fake\texttt{\textless/answer\textgreater}
\end{tcolorbox}

\caption{Speech annotation examples.}
\label{fig:appendix_examples_speech}
\end{figure*}

\begin{figure*}[t]
\centering
\small
\begin{tcolorbox}[
  realexample,
  title={Real Example (Sound)},
  width=\textwidth
]
\textbf{File:} \texttt{EnvSDD/dev/real\_audio/TUTSED2017Dev/a131\_14.wav}\\[1mm]
\texttt{\textless think\textgreater}\\
\textbf{<Frequency Domain>} The audio exhibits a consistent low-frequency hum, characteristic of indoor environmental recordings with nearby electronic equipment. The spectral profile remains stable over time, with no abrupt changes in band energy. No synthetic artifacts (e.g., spectral gaps, unnatural harmonics, or digital clipping) are observed, indicating a smooth and natural frequency response.\\
\textbf{<Time Domain>} The signal maintains a steady background hum without abrupt amplitude changes or unnatural pauses. No transient digital artifacts (clicks/pops) are detected, and the temporal continuity is coherent throughout the clip. The lack of editing traces and the natural stationarity of the noise support the authenticity of the recording.\\
\texttt{\textless/think\textgreater}\\
\texttt{\textless answer\textgreater}real\texttt{\textless/answer\textgreater}
\end{tcolorbox}

\vspace{2mm}

\begin{tcolorbox}[
  fakeexample,
  title={Fake Example (Sound)},
  width=\textwidth
]
{\textbf{File:} \texttt{EnvSDD/dev/fake\_audio/TTA/audioldm1/TUTASC2019Dev/airport-prague-1173-45640-a\_0.wav}}\\[1mm]
\texttt{\textless think\textgreater}\\
\textbf{<Frequency Domain>} The audio shows abnormal high-frequency energy with a metallic timbre and an overly uniform spectral texture, lacking the natural spectral complexity typically present in real-world recordings. Such frequency-domain regularities are consistent with artifacts introduced by audio synthesis models.\\
\textbf{<Time Domain>} The signal exhibits abrupt onset/offset without a natural fade-in/out or reverberant decay. Pitch and amplitude remain unnaturally stable, and background noise lacks realistic fluctuations, indicating weak environmental continuity. These temporal characteristics are typical of AI-generated audio.\\
\texttt{\textless/think\textgreater}\\
\texttt{\textless answer\textgreater}fake\texttt{\textless/answer\textgreater}
\end{tcolorbox}

\caption{Sound annotation examples.}
\label{fig:appendix_examples_sound}
\end{figure*}

\begin{figure*}[t]
\centering
\small
\begin{tcolorbox}[
  realexample,
  title={Real Example (Singing)},
  width=\textwidth
]
\textbf{File:} \texttt{CtrSVDD\_0010\_T\_0008078.flac}\\[1mm]
\texttt{\textless think\textgreater}\\
\textbf{<Frequency Domain>}  The audio exhibits natural vocal harmonics and a smooth spectral distribution without abrupt or synthetic high-frequency cutoffs. Energy is evenly distributed across the frequency range, with no artificial noise patterns typically associated with deepfake generation.\\
\textbf{<Time Domain>} Temporal continuity is preserved throughout the phrase "tenden mushimushi katatsumuri", with seamless transitions between phonemes and consistent breath control. The prosody and timing align with human speech production characteristics, showing no signs of algorithmic manipulation.\\
\texttt{\textless/think\textgreater}\\
\texttt{\textless answer\textgreater}real\texttt{\textless/answer\textgreater}
\end{tcolorbox}

\vspace{2mm}

\begin{tcolorbox}[
  fakeexample,
  title={Fake Example (Singing)},
  width=\textwidth
]
{\textbf{File:} \texttt{CtrSVDD\_0055\_T\_0019326.flac}}\\[1mm]
\texttt{\textless think\textgreater}\\
\textbf{<Frequency Domain>} The audio shows unnatural high-frequency artifacts in sibilant consonants and vowel transitions, with an overly smooth and compressed spectral envelope inconsistent with natural vocal production. A persistent electronic hum further indicates synthetic generation.\\
\textbf{<Time Domain>} Syllable transitions exhibit unnatural timing, particularly in the elongated final syllable \"yuan,\" which lacks human breath control and vibrato modulation. Artificial pauses between syllables and faint static noise align with deepfake synthesis characteristics.\\
\texttt{\textless/think\textgreater}\\
\texttt{\textless answer\textgreater}fake\texttt{\textless/answer\textgreater}
\end{tcolorbox}

\caption{Singing voice annotation examples.}
\label{fig:appendix_examples_sound}
\end{figure*}

\begin{figure*}[t]
\centering
\small
\begin{tcolorbox}[
  realexample,
  title={Real Example (Music)},
  width=\textwidth
]
\textbf{File:} \texttt{lePP2BaZhJU.wav}\\[1mm]
\texttt{\textless think\textgreater}\\
\textbf{<Frequency Domain>}  The audio exhibits a full-frequency range typical of music recordings, with clear and well-defined low, mid, and high frequencies. There are no signs of digital artifacts such as aliasing, quantization noise, or spectral distortion, indicating high fidelity without unnatural frequency imbalances.\\
\textbf{<Time Domain>} The audio maintains a steady rhythm and tempo throughout, with consistent energy across musical segments and no abrupt changes, dropouts, or unnatural silences. The transitions between musical elements are smooth and natural, reflecting authentic performance rather than synthetic generation.\\
\texttt{\textless/think\textgreater}\\
\texttt{\textless answer\textgreater}real\texttt{\textless/answer\textgreater}
\end{tcolorbox}

\vspace{2mm}

\begin{tcolorbox}[
  fakeexample,
  title={Fake Example (Music)},
  width=\textwidth
]
\textbf{File:} \texttt{TTM02\_7jWRIjFaoeU.wav}\\[1mm]
\texttt{\textless think\textgreater}\\
\textbf{<Frequency Domain>} The audio exhibits unnatural high-frequency artifacts characterized by a harsh, clipped texture and absence of natural harmonic overtones, which are indicative of deepfake synthesis rather than genuine acoustic recording. The distorted bass and drums dominate the low-frequency spectrum, while the cymbals and guitar distortion display synthetic high-frequency components inconsistent with organic amplification.\\
\textbf{<Time Domain>} The rhythmic precision and dynamic compression lack variability, suggesting algorithmic generation. Transitions between musical sections occur with mechanical timing, devoid of microtiming nuances inherent to human performance. Additionally, the music ends with an abrupt cutoff lacking natural decay, a common artifact in deepfake audio synthesis.\\
\texttt{\textless/think\textgreater}\\
\texttt{\textless answer\textgreater}fake\texttt{\textless/answer\textgreater}
\end{tcolorbox}

\caption{Music annotation examples.}
\label{fig:appendix_examples_sound}
\end{figure*}

\end{document}